\documentclass[aps,prl,twocolumn,groupedaddress,showpacs]{revtex4}
\usepackage{epsfig}
\begin{document}

\title{NMR-like effect on Anisotropic Magnetic Moment of Surface Bound States in Topological Superfluid $^3$He-B}

\author{M. \v{C}love\v{c}ko}
\author{E. Ga\v{z}o}
\author{M. Skyba}
\author{P. Skyba}
\email[Electronic Address: ]{skyba@saske.sk}

\affiliation{Centre of Low Temperature Physics, Institute  of Experimental Physics SAS,
Watsonova 47, 04001 Ko\v{s}ice, Slovakia.}

\date{\today}

\begin{abstract}

We present experimental observation of a new phenomenon, that we interpret as NMR-like
effect on anisotropic magnetic moment of the surface Andreev bound states in topological
superfluid $^3$He-B at zero temperature limit. We show that an anisotropic magnetic
moment formed near the horizontal surface of a mechanical resonator due to symmetry
violation of the superfluid $^3$He-B order parameter by the resonator's surface may lead
to anomalous damping of the resonator motion in magnetic field. In difference to
classical NMR technique, here NMR was excited using own harmonic motion of the mechanical
resonator, and nuclear magnetic resonance was detected as a maximum in damping when
resonator's angular frequency satisfied the Larmor resonance condition.

\end{abstract}


\maketitle

\section{Introduction}

Superfluid phases of helium-3 provide one of the most complex and  purest physical system
to which we have access to. This unique system is also serving  as a model system for
high energy physics, cosmology and quantum field theories. In fact, the phase transition
of $^3$He into a superfluid state violates simultaneously three symmetries: the orbital,
the spin and the gauge symmetry (SO$^L$(3)$\times$SO$^S$(3)$\times$U(1)). Either A or B
superfluid phase of $^3$He created in zero magnetic field resembles the physical features
comparable with those described by the Standard model or by the Dirac vacuum,
respectively \cite{grisha}. Application of magnetic field breaks the spin symmetry and
this leads to formation of $A_1$ phase in narrow region just below superfluid transition
temperature \cite{A1}. Further, embedding of the anisotropic impurity into $^3$He in form
e.g. nematically ordered aerogel violates the orbital symmetry, which is manifested by
formation of a polar phase of superfluid $^3$He \cite{vovka,jeevak}. Finally, the orbital
symmetry of the superfluid condensate is also violated near the surface of any object of
the size of the coherence length $\xi$ $(\xi \sim 100\,nm)$ being immersed in superfluid
$^3$He-B. Presence of the surface enforces only the superfluid component that consists of
the Cooper pairs having their orbital momenta oriented in direction to the surface normal
and suppresses all others. This results in the distortion of the energy gap in direction
parallel to the surface normal on the distance of a few coherence lengths from the
surface. The gap distortion leads to a strong anisotropy in spin susceptibility of the
superfluid surface layer of $^3$He \cite{nagato2}, as well as to the motional anisotropy
of fermionic excitations trapped in surface Andreev bound states (SABS). It is worth to
note that the dispersion relation of some of these excitations resembles the features of
Majorana fermions, the fermions, which are their own antiparticles
\cite{majorana1,majorana2,majorana3,majorana4}.

This article deals with experimental observations  of a new phenomenon, that we interpret
as NMR-like effect originating from the surface paramagnetic layer in superfluid $^3$He-B
at zero temperature limit. However, in contrast to traditional NMR techniques, the
magnetic resonance was excited using a mechanical resonator oscillating in magnetic field
and detected as an additional, magnetic field dependent mechanical damping of the
resonator's motion.

In order to be able to study physical properties of the surface states using mechanical
resonators, the superfluid $^3$He-B should be cooled to zero temperature limit. When
superfluid $^3$He-B is cooled below 250$\mu$K, a flux of the volume excitations
interacting with a mechanical resonator falls with temperature as $\phi_V\sim
D(p_F)T\exp(-\Delta/k_B T)$ due to presence of the energy gap $\Delta$ in the spectrum of
excitations \cite{andreev}. Here, $D(p_F)$ denotes the density of states at the Fermi
level, $p_F$ is the Fermi momentum and $k_B$ is the Boltzmann constant. On the other
hand, the gap distortion in vicinity of the surface modifies the dispersion relation for
the excitations trapped in SABS to "A-like" phase and corresponding flux of the surface
excitations $\phi_S$ varies non-exponentially $(\phi_S \sim D_S(p_F)T^3)$, where
$D_S(p_F)$ is the density of states near surface which depends on the surface quality.
Therefore, one may expect that in superfluid $^3$He-B at higher temperatures
$\phi_S<\phi_V$ (in a hydrodynamic regime), while at ultra low temperatures (in a
ballistic regime) a state when  $\phi_S>\phi_V$ can be achieved. This temperature
transition can be detected using e.g. mechanical resonators as a decrease of their
sensitivity to the collisions with volume excitations with temperature drop. It is
obvious that this transition temperature depends on the resonator's mass, the area and
quality of the resonator's surface which determines the density of the surface states
\cite{nagato2,nagato1,sauls,aoki}.

\begin{figure}[htb]
\begin{center}
\epsfig{figure=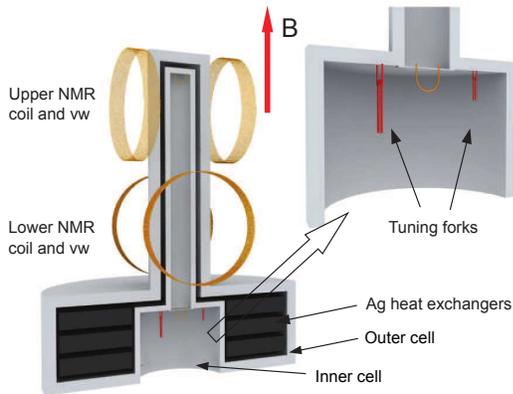,width=80mm} \caption{(Color online) A schematic sketch of the double
walled experimental cell mounted on our nuclear stage. The orientation of magnetic field $B$
is shown as well.  }\label{fig1}
\end{center}
\end{figure}

There are variety of mechanical resonators being used as experimental tools to probe the
physics of topological $^3$He-B at zero temperature limit
\cite{bradley0,bradley1,zheng1,zheng2}. As  mechanical resonators we utilize tuning
forks. Currently, these piezoelectric devices are very popular experimental tools used in
superfluid $^3$He physics \cite{rob}. They are almost magnetic field insensitive, simple
to install, easy to excite with extremely low dissipation  of the order of a few fW or
even less and displacement $\sim$ 0.1nm and straight forward to measure. The measured
current $I_F$ is proportional to the fork velocity $I_F = A v$, where $A$ is the
proportionality constant readily determined from experiment \cite{skybos,IV}.

\section{Experimental details}

We performed experiments in a double walled experimental cell (see Fig. \ref{fig1})
mounted on a diffusion-welded copper nuclear stage \cite{stage}.  While upper tower
served for NMR measurements (not mentioned here), in the lower part of the experimental
cell,  tuning forks of different sizes and one NbTi vibrating wire were mounted. The NbTi
vibrating wire served as a thermometer of $^3$He-B in ballistic regime i.e. in
temperature range below 250$\mu$K.

After cooling the fridge down to $\sim$ 0.9\,K, we initially characterized the tuning
forks in vacuum using a standard frequency sweep technique in order to determine $A$
constants for the individual forks \cite{rob,skybos}. Both forks behaved as high Q-value
resonators having Q-value of the order of 10$^6$. The physical characteristics of the
large and small tuning forks are as follow:  the large fork resonance frequency (in
vacuum) $f_0^L$ = 32725.88\,Hz, the width $\Delta f_{2i}$ = 36.3\,mHz and dimensions
$L=3.12$\,mm, $W=0.25$\,mm,  $T=0.402$\,mm give the mass m$_{L}$=2.0$\times$10$^{-7}$\,kg
and value of $A_{L}$ = 6.26$\times$10$^{-6}$ A.s.m$^{-1}$, while the small fork resonance
frequency $f_0^S$ = 32712.968\,Hz, the width $\Delta f_{2i}$ = 32.79\,mHz and dimensions
$L=1.625$\,mm, $W=0.1$\,mm, $T=0.1$\,mm give the mass m$_S$=1.05$\times$ 10$^{-8}$\,kg
and the value of $A_S$ = 1.04$\times$10$^{-6}$ A.s.m$^{-1}$.

Then, we filled the experimental cell with $^3$He at pressure of 0.1\,bar. Subsequent
demagnetizations of the copper nuclear stage allowed us to cool the superfluid $^3$He-B in
the inner cell down to 175$\mu$K as determined from the damping of the NbTi vibrating
wire.

We also performed measurements of tuning forks in small magnetic fields. After
demagnetization, when temperature of the superfluid $^3$He-B was stable, we set the
magnetic field $(B_0)$ to 2.5\,mT, and measured  a collection of the resonance curves at
various excitations at this field. Thereafter, we reduced the magnetic field $B_0$ slowly
by 0.25\,mT and repeated the measurements of the resonance characteristics as a function
of excitation. We reproduced this measurement procedure while reducing the magnetic field
$B_0$ down to 0.25\,mT.

\begin{figure}[htb]
\begin{center}
\epsfig{figure=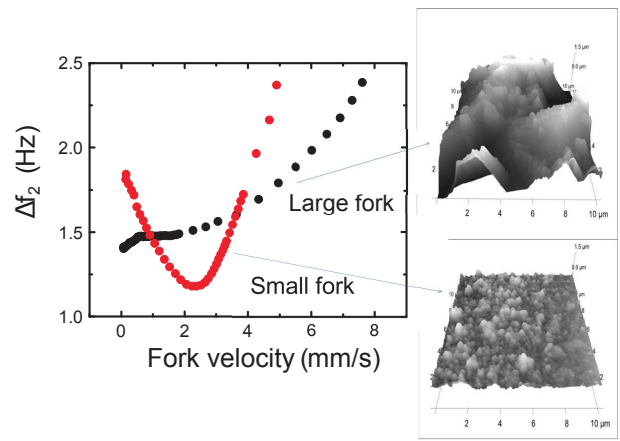,width=80mm} \caption{(Color online) Dependence of the width $\Delta f_2$
for large (black) and small (red) tuning forks as a function of fork velocity and
images of the surface profiles of tuning forks used in experiment obtained from AFM scans.} \label{fig2}
\end{center}
\end{figure}

\section{Experimental results and analysis}

Figure \ref{fig2} shows the width $\Delta f_2$ as a function of the fork ve\-lo\-city
measured for the large and small tuning forks in superfluid $^3$He-B at temperature of
175\,$\mu$K and pressure of 0.1\,bar. As one can see, there is a remarkable difference
between them. The small tuning fork clearly demonstrates  Andreev reflection process
\cite{andreev,bradley,jltp}: as the fork velocity is rising more and more volume
excitations undergo the process of  Andreev reflection.  During this scattering process
they exchange a tiny momentum with the fork of the order of $(\Delta /E_F) p_F$, where
$E_F$ and $p_F$ are the Fermi energy and the Fermi momentum, respectively. As a result,
the fork damping decreases until a critical velocity is reached. At this velocity the
fork begins to break the Cooper pairs and its damping rises again. However, this
dependence for the large tuning fork is opposite: the process of the Andreev reflection
is suppressed, and the damping of the large fork increases with its velocity at the
beginning. The different width $\Delta f_2$ - velocity dependence for the large fork
presented in Fig. \ref{fig2} suggests a presence of some processes leading to the
suppression of the Andreev reflection and/or another dissipation mechanism than the
Andreev reflection.

\begin{figure}[htb]
\begin{center}
\epsfig{figure=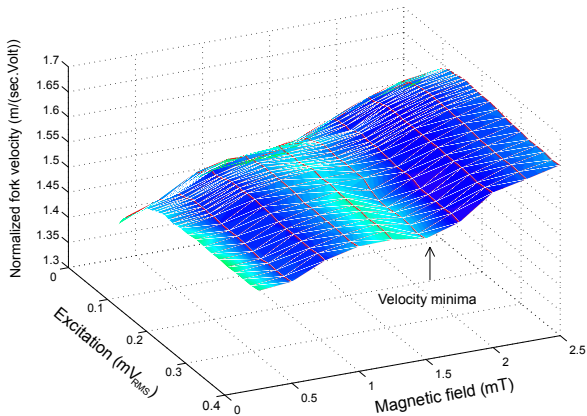,width=90mm} \caption{(Color online) Dependence of the
normalized tuning fork velocity (fork velocity over excitation voltage) as a function of the fork
velocity and applied magnetic field $B_0$. Dependence clearly shows a presence of the
velocity minima at magnetic fields that satisfy the Larmor resonance condition
$\omega_L = \gamma (B_0 + B_{rem})$. } \label{fig3}
\end{center}
\end{figure}

We assume that at temperature $\sim 175 \mu$K, the density of excitations near the
surface of large fork satisfies the condition $\phi_S > \phi_V$. However, we suppose that
``non-standard'' behavior of the large tuning fork is caused by the different quality of
its surface compared with that of the small fork. While the surface of the large fork is
corrugated, the surface of the small fork is much smoother (see Fig. \ref{fig2}). Fork
motion in superfluid $^3$He-B is associated with creation of the back-flow i.e. the flow
of the superfluid component around tuning fork's body on the scale of the slip length
\cite{slip1,slip2}. The back-flow shifts the energy of excitations by ${\mathbf p}_F
\cdot \mathbf{v}$, where ${\mathbf v}$ is the superfluid velocity (in linear
approximation is the same as the fork velocity). As a consequence, the excitations having
energy less than $\Delta + {\mathbf p}_F\cdot {\mathbf v}$ are scattered via Andreev
process. This simple model assumes that direction of the back-flow is correlated with the
direction of the fork velocity i.e. the back-flow flows in opposite direction to the
tuning fork motion. However, assuming that the scale of the surface roughness of the
large fork is larger than the slip length (see Fig. \ref{fig2}), the oscillating surface
of the large fork makes the velocity field of the superfluid back-flow random. That is,
the back-flow is not correlated with direction of the tuning fork motion. This means that
there are excitations reflected via Andreev process there due to the back-flow flowing in
different directions to that of the tuning fork velocity. Such reflected excitations are
practically ``invisible'' to the fork. On the other hand, the small fork is sensitive to
the Andreev reflection supposing that the scale of its surface roughness is less or
comparable to the slip length. We presume that above presented mechanism stays behind the
suppression of the Andreev reflection in case of the large fork. However, to confirm this
hypothesis additional work has to be done. Regarding to the rise of the large fork
damping at low velocities, the origin of this phenomenon is unclear yet, and it is not a
subject of this article.

Another unexpected results were observed while measuring the damping of the large tuning
fork motion in superfluid $^3$He-B at temperature of 175$\mu$K in magnetic field. We
surprisingly found that the damping of the large fork motion is magnetic field dependent
and shows a maximum.

\begin{figure}[htb]
\begin{center}
\epsfig{figure=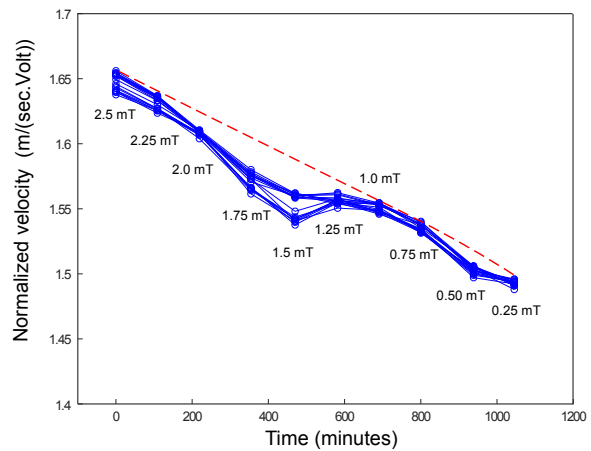,width=85mm} \caption{(Color online) Time dependencies of the normalized
tuning fork velocity (fork velocity over excitation voltage) measured at different magnetic fields
as showed. The points represent the data measured for various excitations at particular field.
Figure clearly shows a presence of the velocity minima at magnetic field corresponding to the
Larmor resonance condition $\omega_L = \gamma (B_0 + B_{rem})$. The dashed line illustrates a small
thermal background caused by a parasitic heat leak.} \label{fig4}
\end{center}
\end{figure}

\begin{figure}[htb]
\begin{center}
\epsfig{figure=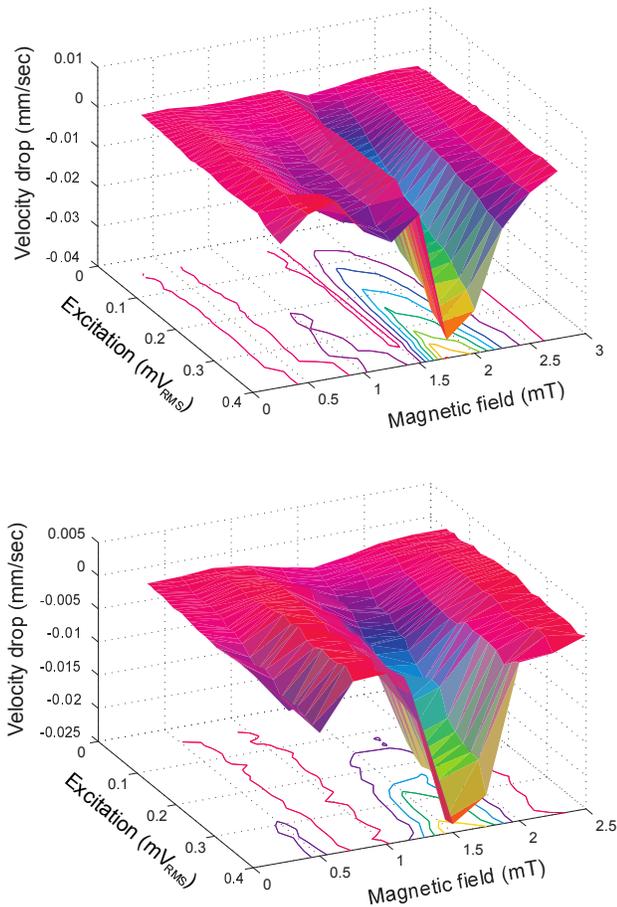,width=85mm} \caption{(Color online) Two dependencies of the
velocity drop of the tuning fork expressed as a function of the magnetic field $B_0$ and
excitation voltage measured during two subsequent demagnetizations. Both dependencies show
a clear minimum in velocity as function of the magnetic field at value that corresponds
to the Larmor resonance frequency $\omega_F = \gamma (B_0+B_{rem})$. We note that value
of $B_{rem}$ is $\sim$  -0.5\,mT.}
\label{fig5}
\end{center}
\end{figure}

Figure \ref{fig3} shows the results of above mentioned measurements in a form of the
dependence of the normalized tuning fork velocity as a function of the fork velocity and
magnetic field. This dependence clearly shows a presence of the minima in the fork
velocity at the same value of the magnetic field. Presented dependencies are masked by a
tiny thermal background due to a small warm-up caused by a parasitic heat leak into
nuclear stage (see Fig. \ref{fig4}). Time evolution of the thermal background was modeled
using the polynomial dependence $a\cdot t^2+b\cdot t+c$, where $a,b,c$ are fitting
parameters and $t$ is the time. We determined these parameters for particular excitation
by fitting the time dependence via points measured at 2.5, 2.25, 1.0, 0.75 and 0.25\,mT
(see illustrative the red dashed line in Fig. \ref{fig4}). When we subtracted-off the
thermal background, the resulting dependencies are presented in Fig. \ref{fig5}. Figure
\ref{fig5} shows two dependencies of the tuning fork velocity as a function of excitation
and magnetic field $B_0$.  These two dependencies measured during two subsequent
demagnetizations demonstrate their reproducibility.

Figure \ref{fig5} manifests a new and intriguing phenomenon: a presence of the velocity
minima (i.e. an additional damping) at magnetic fields which satisfy the Larmor resonance
condition $\omega = \gamma (B_0 + B_{rem}) = \gamma B$, where $\omega$ is the angular
frequency of the tuning fork ($\omega \simeq 2\pi \cdot$32.4\,kHz), $\gamma$ is the
$^3$He gyromagnetic ratio $(\gamma = - 2\pi\cdot 32.4 \times 10^3\,rad/s.mT)$, $B_0$ and
$B_{rem}$ is the magnetic field applied and remnant magnetic field from demagnetization
magnet, respectively. We interpret this phenomenon as NMR-like effect on the anisotropic
magnetic moment $\mathbf M$ formed in vicinity of the top horizontal surface of the
tuning fork. Formation of the anisotropic magnetic moment $\mathbf M$ is a consequence of
the symmetry violation of the $^3$He-B order parameter by the fork surface,
simultaneously modifying the excitation spectrum. Based on different behaviors of the
tuning forks showed in Fig. \ref{fig2} we presume that the damping of the large tuning
fork motion is mostly caused by the excitations trapped in the surface states, as the
rest of superfluid $^3$He-B in volume behaves like a vacuum.

\begin{figure}[tb]
\begin{center}
\epsfig{figure=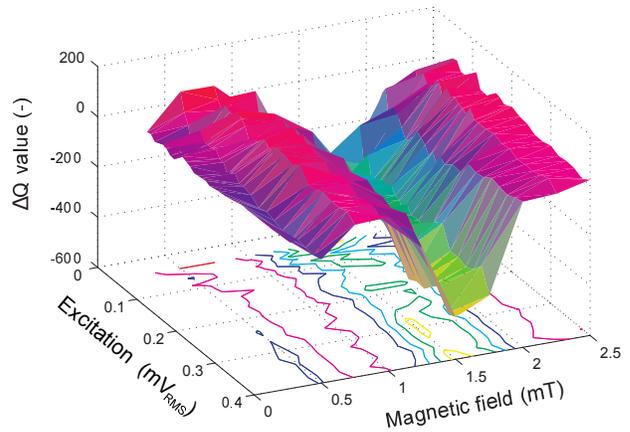,width=85mm} \caption{(Color online) Dependence of the tuning
fork Q-values as a function of magnetic field $B_0$ and excitation voltage. The
deep minimum corresponds to nuclear magnetic resonance of the magnetic layer formed
on tuning fork surface.}
\label{fig6}
\end{center}
\end{figure}

Figure \ref{fig6} shows the same NMR effect, however, as a drop of the tuning fork
Q-value in dependence on the excitation and magnetic field.

In order to explain the measured dependencies we propose a simple phenomenological model
as follows. The fork's motion in zero magnetic field can be described by the equation
\begin{equation}
\frac{d^2 \alpha}{d t^2} +  \Gamma \frac{d \alpha}{d t} + \omega_0 \alpha = F_m \sin (\omega t)\,, \label{equ1}
\end{equation}
where $\Gamma$ is the damping coefficient characterizing the fork's interaction with
surrounding superfluid $^3$He-B, and which also includes its own intrinsic damping,
$\omega_0$ is the fork resonance frequency in vacuum, $\omega$ is the angular frequency
of the external force, $\alpha$ is the deflection angle of the tuning fork arm from
equilibrium and $F_m$ is the force amplitude normalized by the mass $m$ and by the length
$l$ of the tuning fork arm. We assume that fork's deflections are small enough and
therefore interaction of the tuning fork with superfluid $^3$He-B acts in linear regime
i.e. we neglect processes of the Andreev reflection \cite{andreev,jltp}.

By applying external magnetic field $\mathbf{B}=(0,0,B_0 + B_{rem})=(0,0,B)$, the
magnetic moment  $\mathbf{M}_0$ is formed on the fork's horizontal surfaces. The magnetic
moment ${\mathbf M}_0$ of the surface layer includes the strong spin anisotropy of the
superfluid $^3$He layer together with magnetic moments of solid $^3$He atoms covering the
fork's surface \cite{nagato2,prl2010,silaev1,silaev2,silaev3}. However, based on
measurements presented in \cite{prl2010}, we presume that magnetic moments of solid
$^3$He atoms behave as a paramagnet and, on a time scale of the fork oscillation period,
its Zeeman energy is always minimized. Therefore, magnetic property of the surface layer
of superfluid $^3$He-B is responsible for the anisotropy of the magnetic moment ${\mathbf
M}$. According to \cite{nagato2}, the anisotropic spin susceptibility of the surface
layer of superfluid $^3$He-B at T$\rightarrow$0 can be expressed as
\begin{equation}
\chi_{zz}=\frac{\hbar^2 \gamma^2 k_F^2}{16 \pi \Delta} \,,
\end{equation}
where $p_F=\hbar k_F$. This susceptibility is as large as the normal state susceptibility
$\chi_N$ multiplied by the width $1/\kappa = \hbar\,v_F/\Delta$ of the bound states. Here, $v_F$
is the Fermi velocity.

\begin{figure}[t]
\begin{center}
\epsfig{figure=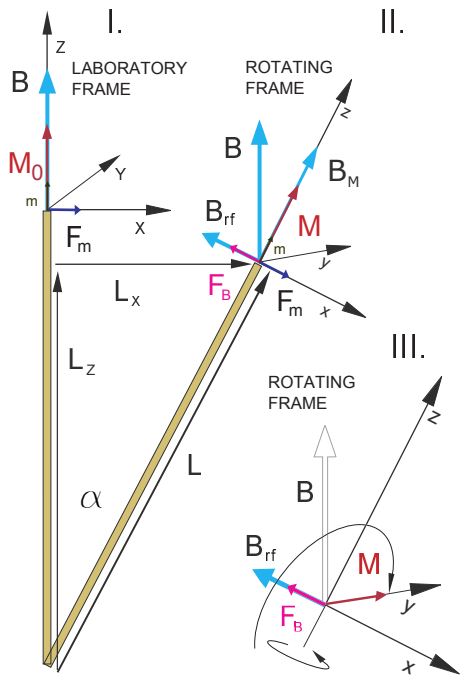,width=60mm} \caption{(Color online) Schematic view on individual
vectors for two positions of the tuning fork. In stationary position (I.) Zeeman energy is
minimized and magnetic torque is zero. When fork is deflected by external force $\mathbf{F}_m$
(II.), the rise of Zeeman energy due to anisotropy of the magnetic moment $\mathbf{M}$ is
associated with emergence of the force $\mathbf{F}_B$ which acts against external force
$\mathbf{F}_m$ and causes additional, magnetic field dependent damping. When NMR condition is
satisfied (III.) i.e. when $B_{eff}=B_{rf}$, magnetic moment $\mathbf{M}$ precesses in $z-y$
plane around $B_{rf}$ in rotating frame of the reference.} \label{fig7}
\end{center}
\end{figure}

In general, due to surface diffusivity the orientation of ${\mathbf M}_0$ can be tilted
from the field direction ${\mathbf B}$.  However, we shall assume for simplicity that
${\mathbf M}_0 = (0,0,M_0)$. When fork oscillates in external magnetic field
$\mathbf{B}$, the normal to its horizontal surface $\mathbf m$ is deflected  from the
direction of the external magnetic field $\mathbf{B}$ (see Fig.\,\ref{fig7}). This means
that anisotropic magnetic moment $\mathbf{M}$ of the surface layer undergoes the same
deflections (oscillations). We assume that during fork oscillations the magnetic moment
of solid $^3$He layer follows the direction of magnetic field ${\mathbf B}$ minimizing
its Zeeman energy. Therefore, in the reference frame connected to the anisotropic
magnetic moment $\mathbf{M}$, this moment $\mathbf{M}$ experiences a linearly polarized
alternating magnetic field $\mathbf{B}_{rf}$ of amplitude $B_{rf}=B \sin (\alpha)\simeq B
\alpha$ oscillating with angular frequency $\omega$  of the tuning fork. While magnetic
field magnitude in the direction of magnetic moment $\mathbf{M}$ is $B_M = B \cos
(\alpha)\simeq B$. Thus, a typical experimental NMR configuration is set up. However,
here the ``virtual'' excitation rf-field $\mathbf{B}_{rf}$ acting on anisotropic magnetic
moment $\mathbf{M}$ is generated by harmonic mechanical motion of the tuning fork. We
suppose that magnetic torque $\mathbf{M} \times \mathbf{B}$ acting on the anisotropic
magnetic moment $\mathbf{M}$ is equivalent to the mechanical torque $\mathbf{L} \times
\mathbf{F}_B$, where $\mathbf{L}=(L_X,0,L_Z)$ is the vector pointed in direction of
$\mathbf{M}$ having the magnitude equal to the length of the oscillating fork prong $l$.
The force $\mathbf{F}_B$ emerges from the rising of the Zeeman energy
($-\mathbf{M}.\mathbf{B})$ due to deflection of $\mathbf{M}$ from the field direction,
acts against excitation force $\mathbf{F}_m$ and causes additional field-dependent
damping of the tuning fork motion. This force can be expressed as
\begin{equation}\label{equ2}
{\mathbf F_B} = \frac{1}{\gamma l^2}\frac{d {\mathbf M}}{d t} \times {\mathbf L}\,.
\end{equation}
In order to obtain time dependence of $\mathbf{F}_B$ one has to determine dynamics of the
magnetic moment $\mathbf{M}$ which is governed by the Bloch's equation (in rotating
reference frame)
\begin{equation}\label{equ3}
\frac{\delta \mathbf{M}}{\delta t} =   \gamma \left(\mathbf{M} \times \mathbf{B}_{eff}\right)
+ \frac{\mathbf{M}_0 - \mathbf{M}}{T_i}\,.
\end{equation}
Here $\mathbf{M} =(M_x,M_y,M_z)$, $\mathbf{B}_{eff} = (-B_{rf}, 0, B - \omega/\gamma)$,
and the second term on the right side describes the processes of the energy dissipation
being characterized by the relaxation time constants $T_i$ with $i=1$ for $z$-component
and $i=2$ for $xy$-component of the magnetic moment~$\mathbf{M}$. Assuming that magnetic
relaxation processes act solely in the magnetic layer near fork surface and using the
geometry of the problem, the amplitude of the force $F_B$ acting against excitation force
$F_m$ can be expressed as
\begin{equation}\label{equ4}
F_B = \frac{1}{\gamma l}\frac{d M_Y}{d t} = \frac{1}{\gamma l } \left[\chi_D \cos(\omega t)+
\chi_A \sin (\omega t)\right] B \frac{d
\alpha}{ d t}\,,
\end{equation}
where $M_Y$ is the y-component of magnetic moment $\mathbf{M}$ in the laboratory frame,
$\omega$ is the tuning fork angular frequency, $\chi_A$ denotes the absorption component
of the magnetic susceptibility of the layer  expressed in the form
\begin{equation}\label{equ5}
\chi_A =\frac{\chi_L \omega_B T_2}{1 + (\omega_B-\omega)^2T_2^2}
\end{equation}
and $\chi_D$ denotes the dispersion component in the form
\begin{equation}\label{equ6}
\chi_D =\frac{\chi_L \omega_B (\omega_B-\omega)T_2^2}{1 + (\omega_B-\omega)^2T_2^2}\,.
\end{equation}
Here $\omega_B = \gamma B$ and $M =\chi_L B$, where $\chi_L$ stands for the magnetic
susceptibility of the $^3$He-layer. Adding the term (\ref{equ4}) into equation
(\ref{equ1}), one gets nonlinear differential equation describing the tuning fork motion
with anisotropic magnetic layer on its horizontal surface in external magnetic field in
form

\begin{equation}
\frac{d^2 \alpha}{d t^2} + \Gamma \frac{d \alpha}{d t} + \frac{1}{\gamma l^2 m}\frac{d M_Y}{d t}+
\omega_0 \alpha = F_m \sin (\omega t)\,.
\label{equ7}
\end{equation}

\begin{figure}[tb]
\begin{center}
\epsfig{figure=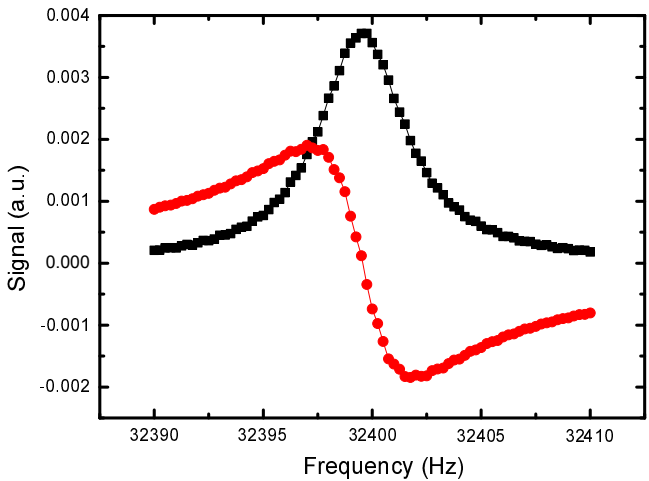,width=85mm} \caption{(Color online)
Example of the calculated resonance characteristics of the tuning fork using equation
(\ref{equ7}) for magnetic field 2.25\,mT and excitation 175\,mV.}
\label{fig8}
\end{center}
\end{figure}

\begin{figure}[htb]
\begin{center}
\epsfig{figure=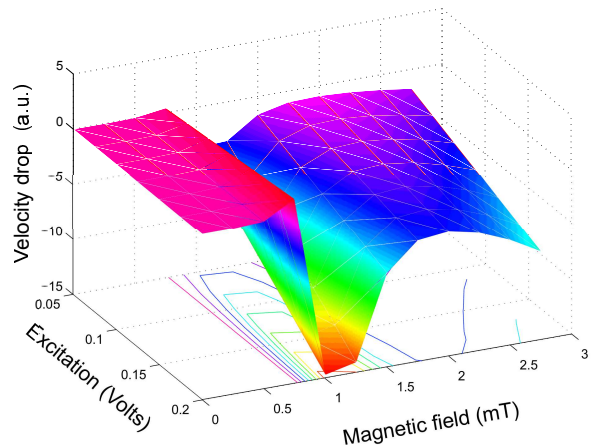,width=80mm} \caption{(Color online) Theoretical dependence of the
tuning fork velocity drop as a function of the magnetic field $B_0$ and excitation voltage
calculated using equation (\ref{equ7}).} \label{fig9}
\end{center}
\end{figure}

Applying Runge-Kutta method we  numerically calculated the time evolution of the tuning
fork response described by the equation (\ref{equ7}) as a function of applied external
force (in frequency and amplitude) and magnetic field. Calculations took into account
transient phenomena. Reaching a steady state of the fork motion, the calculated values
were multiplied by the "reference" signals simulating excitations with aim to obtain the
resonance characteristics i.e. the absorption and the dispersion component. The magnetic
properties of surface layer were characterized by the spin-spin relaxation time constant
$T_2$=28\,$\mu$sec which served as a fitting parameter. Figure \ref{fig8} shows an
example of the tuning fork response calculated by using equation (\ref{equ7}) in form of
the resonance characteristics. Calculated resonance characteristics were fitted by means
of the Lorentz function in order to obtain experimentally measurable parameters: the
velocity amplitude, the width and the resonance frequency as the function of the
excitation and magnetic field. Figure \ref{fig9} summarizes theoretically calculated
dependence of the tuning fork velocity drop as a function of driving force (excitation
voltage) and magnetic field. Presented dependencies confirm the presence of the velocity
minima at magnetic field corresponding to the Larmor resonance condition for $^3$He and
they are in very good qualitative agreement with those obtained experimentally (see Fig.
\ref{fig5}).

\section{Discussion}

Although, we have presented a simple theoretical model using a phenomenological approach,
we obtained  reasonable qualitative agreement with experiment. However, it is worth to
say that there is a set of theoretical papers dealing with problem of the
Andreev-Majorana surface states in topological superfluid $^3$He-B on the level of the
order parameter
\cite{nagato2,volovik1,silaev1,silaev2,silaev3,sauls1,tsutsumi,takeshi,takeshi1}. In
light of this, let us discuss our experimental results and compare them with theoretical
models.

In particular, the spin dynamics and an effect of NMR  on the magnetic moment of the
surface states had recently been theoretically investigated by M. A. Silaev
\cite{silaev3}. Using assumptions of a flat surface, i.e. that the vector ${\mathbf n}$
representing a rotation axis of the superfluid $^3$He-B order parameter is parallel to
magnetic field ${\mathbf B}_0$, he showed that standard transverse NMR technique does not
allow to excite the magnetic moment of the surface states at Larmor frequency due to two
reasons. The first, a mini-gap presented at the surface state spectrum has a broader
energy gap $E_g$ than corresponding Larmor frequency due to Fermi-liquid corrections
($E_g \sim 4 \hbar \omega_L$). The second, the probability of the NMR excitation in
surface states spectrum is proportional to the deflection angle $\beta_n$ of the vector
${\mathbf n}$ from the spin quantization axis defined by magnetic field ${\mathbf B}_0$.
For a flat surface, the angle $\beta_n$ is rather small leading to a strong suppression
of the NMR response from the surface states. However, in order to excite the
Andreev-Majorana surface states using a transverse NMR technique, the vector ${\mathbf
n}$ has to be deflected from magnetic field direction by an angle $\beta_n$, so that the
effective driving field ${\mathbf B}_{eff}$ has a component parallel to the spin
quantization axis \cite{silaev3}.

The Lancaster group \cite{fisher} performed the NMR  measurements using superfluid
$^3$He-B near a surface in experimental configuration, which is  similar to that as
assumed in \cite{silaev3}. The experimental cell was made from sapphire. However, instead
of a flat horizontal surface, their cell had a semi-spherical end cap. Semi-spherical end
cap of the experimental cell formed a texture of $\mathbf{n}$-vectors in broad range of
angles $\beta_n$ with respect to the direction of magnetic field ({\it z}-direction).
This is a configuration for which the theory \cite{silaev3} predicts possibility to
excite and observe response from the Andreev-Majorana surface states. Pulsed NMR
technique allowed them to create a long lived state with coherent spin precession named
as persistently precessing domain (PPD) \cite{ppd,kupka}. Using magnetic field gradient
they were able to control the position of the PPD with respect to the horizontal wall of
the cell \cite{spatial}. They showed that the closer the Larmor resonance condition is to
the cell horizontal surface, the shorter the PPD signal life time is. The presence of the
cell surface reduces the signal life time by four orders of magnitude \cite{fisher}. In
the light of theory \cite{silaev3}, the interpretation of this phenomena needs to be
elucidated.

Our experimental configuration of the NMR detection using mechanical resonator is
completely different from that of a standard transverse NMR technique. The most important
difference is that we did not apply any external rf-field ${\mathbf B_{rf}}$ to excite
magnetic moments in superfluid $^3$He-B. The only magnetic field presented is the static
magnetic field ${\mathbf B}$. An excitation rf-field ${\mathbf B_{rf}}$ is a ``virtual''
field, which is experienced only by the anisotropic magnetic moment ${\mathbf M}$ during
fork oscillations.

Harmonic oscillations of the tuning fork arms lead to the oscillation of the whole
texture of ${\mathbf n}$ vectors causing their time dependent deflection from the
quantized axis defined by the constant magnetic field ${\mathbf B}$. The amplitude of the
tuning fork oscillation at maximum excitation was $\sim$ 2\,nm. This is much  less than
the coherence length and the size of surface roughness. Low oscillation amplitude also
reduces the bulk effects. Diffusivity of the tuning fork surface (see Fig. \ref{fig2})
ensures the deflections of the ${\mathbf n}$ vectors in a broad spectrum of angles
$\beta_n$ from the field direction. This is a configuration, which according to model
\cite{silaev3}, satisfies the condition for the observation of transverse NMR of the
surface states in superfluid $^3$He-B. However, according to our opinion, the assumptions
of the theoretical model \cite{silaev3} do not fully correspond to the conditions of the
experiment presented here and the model itself could be extended for them.

On the other hand, theoretical model presented in \cite{nagato2}  suggests that strong
spin anisotropy of superfluid $^3$He layer near the surface is large enough to be
observed experimentally. We think that above mentioned experimental technique using
mechanical resonators (e.g. tuning forks) with various surface roughness and resonance
frequencies is a way. However, the open question is an influence of the solid $^3$He on
this phenomenon. Although we assumed a paramagnetic property of the solid $^3$He layer,
the magnetic susceptibility of the solid $^3$He dominates at ultra low temperatures
\cite{prl2010}. An influence of solid $^3$He could be tested by using $^4$He as a
coverage on the surface of the tuning fork, since $^4$He atoms remove $^3$He atoms from
the surface. However, $^4$He simultaneously covers the heat exchangers inside nuclear
stage and this reduces the cooling efficiency of the $^3$He liquid. As the measurements
are performed in temperature range below 200\,$\mu$K, the test of influence of the $^3$He
solid layer on the observed phenomenon is going to be an experimental challenge.

\section{Conclusion}

In conclusion, we have observed the NMR-like effect on anisotropic magnetic moment of the
superfluid surface layer, including Andreev-Majorana fermionic excitations formed on the
resonators' surface, being detected as additional magnetic field dependent damping of its
mechanical motion. Further work is required to develop this technique, which in
combination with e.g. acoustic method \cite{majorana2}, or non-standard NMR technique
based on PPD \cite{fisher}, or thermodynamic method \cite{bunkov} opens a possibility to
study the spin dynamics of excitations trapped in SABS in topological superfluid $^3$He-B
and, perhaps to prove their Majorana character experimentally. Finally, a development of
a theory considering oscillations of ${\mathbf n}$ vectors representing the order
parameter of superfluid $^3$He near a surface at constant magnetic field, i.e. a theory
for the condition of presented experiment would be very useful and challenging.

\section{Acknowledgments}

We acknowledge the support of APVV-14-0605, VEGA 2/0157/14,  Extrem II - ITMS 26220120047
and European Microkelvin Platform, H2020 project 824109. We wish to thank to M. Kupka for
fruitful discussion, V. Komanick\'y for AFM scans of tuning forks, and to \v{S}. Bic\'ak
and G. Prist\'a\v{s} for technical support. Support provided by the U.S. Steel Ko\v{s}ice
s.r.o. is also very appreciated.

\end{document}